\documentclass[sigconf]{acmart}
\AtBeginDocument{%
  \providecommand\BibTeX{{%
    \normalfont B\kern-0.5em{\scshape i\kern-0.25em b}\kern-0.8em\TeX}}}

\newcommand{\reffig}[1]{Figure~\ref{#1}}
\hypersetup{pdfstartview={XYZ null null 1.00}}
\begin{document}

\title{Just Enough, Just in Time, Just for ``Me'': Fundamental Principles for Engineering IoT-native Software Systems}


\author{Zheng Li}
\affiliation{%
  \institution{University of Concepci{\'o}n}
  \city{Concepci{\'o}n}
  \country{Chile}}
\email{imlizheng@gmail.com}

\author{Rajiv Ranjan}
\affiliation{%
  \institution{Newcastle University}
  \city{Newcastle upon Tyne}
  \country{UK}}
\email{raj.ranjan@ncl.ac.uk}
\begin{abstract}
By seamlessly integrating everyday objects and by changing the way we interact with our surroundings, Internet of Things (IoT) is drastically improving the life quality of households and enhancing the productivity of businesses. Given the unique IoT characteristics, IoT applications have emerged distinctively from the mainstream application types. Inspired by the outlook of a programmable world, we further foresee an IoT-native trend in designing, developing, deploying, and maintaining software systems.
However, although the challenges of IoT software projects are frequently discussed, addressing those challenges are still in the ``crossing the chasm'' period. By participating in a few various IoT projects, we gradually distilled three fundamental principles for engineering IoT-native software systems, such as \textit{just enough}, \textit{just in time}, and \textit{just for ``me''}. These principles target the challenges that are associated with the most typical features of IoT environments, ranging from resource limits to technology heterogeneity of IoT devices. We expect this research to trigger dedicated efforts, techniques and theories for the topic IoT-native software engineering.
 
\end{abstract}



\maketitle

\section{Introduction}
Driven by the revolutionary development of modern computer, information and communication technologies (e.g., from large mainframes to small chips and from coaxial cables to 5G), all the sectors of our lives are experiencing a rapid evolution in the use of Internet of Things (IoT) \cite{Caro_Navarro_2018,Khanna_2019,Wang_Lim_2021}. An estimation shows that about 50 billion devices will be connected and in use worldwide by 2030 for creating smart environments \cite{TheStar_2021}, while the recent statistics reveal that ``an increase of 10 percentage points in the growth of IoT connections per inhabitant is associated with a 0.23 percentage points increase in the total factor productivity (TFP) growth'' \cite{Eduist_2021}. Correspondingly, large amounts of companies have started investing on IoT applications to grow their businesses and to strengthen their competitive advantages.

Despite the de facto cloud-native solutions that satisfy IoT application requirements mainly in the data centers (e.g., AWS IoT and Azure IoT), there is a clear trend in deploying and executing custom application logics directly on IoT devices \cite{Vogler_2016}. In fact, by foreseeing richer embedded execution environments, researchers have predicted that our everyday things will all become not only connected but also programmable dynamically \cite{Taivalsaari_2017}. Inspired by the prospect of a programmable (Internet) edge world, we advocate IoT-native solutions for achieving IoT-intensive business values, by composing distributed software components across IoT and edge nodes. In specific, we define IoT-native software to be executable artefacts designed, developed and deployed directly in the IoT environments, with little or limited cloud-side orchestration.

However, there are inevitable challenges of engineering IoT-native software systems, and the challenges are closely associated with the IoT characteristics. In addition to the massive device distribution \cite{Morin_2017} that leads to the generic challenges of distributed software engineering \cite{Rotem-Gal-Oz_2013}, this research particularly highlights two fundamental features of IoT.
First, \textbf{individual IoT devices generally have limited compute, storage and network capabilities}. For example, the embedded RAM is only 4KB for current taximounted GPS devices. Although the optimistic opinion about future IoT hardware is that any type of device will enable virtualized software stacks (ranging from fully fledged OSs to dynamic language runtimes) \cite{Taivalsaari_2017}, those everyday things shall always have considerably less resources than the dedicated computing machines of the same generation.

Second, \textbf{an IoT system may involve non-scalable integration of heterogeneous technologies produced by different manufacturers}. On one hand, the IoT industry has not employed any globally standard protocols and universal interfaces yet, which is similar to the early computer industry \cite{Gorman_Resseguie_2009}. On the other hand, to secure the niche market shares against competitors, different device vendors even tend to keep developing their own proprietary standards and continue the existing incompatible protocols. Although virtualization and abstraction are the promising strategies to standardize the programming and execution environment \cite{Zambonelli_2017}, they are still far from the current reality of ``things''. Furthermore, to take advantage of manufacturer-specific technologies, the heterogeneity may need to remain explicit and exploitable by IoT application developers \cite{Morin_2017}.

To address or at least relieve these challenges, we gradually summarized our experience and lessons from a few various IoT software projects into three fundamental principles, such as \textit{just enough}, \textit{just in time}, and \textit{just for ``me''}. In brief, (1) \textit{just enough} aims to address the limitation of compute and storage capabilities of IoT devices, by minimizing data collection, storage duration, and processing workloads. (2) \textit{Just in time} aims to address IoT devices' limited network capability and/or the environmental bandwidth constraints, by optimizing the data communication among IoT application components. (3) \textit{Just for ``me''} aims to address the heterogeneity and complexity in organizing the distributed IoT application components, by enabling the components' awareness of environments and by making them collaborate with each other without using orchestration flows. More importantly, in this paper we not only justify the three principles theoretically, but also demonstrate suitable techniques that can satisfy these principles, based on our practices in IoT projects \cite{Li_Pino_2019,Li_Seco_2021,Li_Pedro_2022}. As such, this research makes a twofold contribution:

\begin{itemize}
\item In theory, our forward-looking vision about IoT-native software can inspire researchers to keep developing a dedicated suite of software engineering principles for IoT. In fact, it has been identified that IoT applications differ from the mainstream application types \cite{Taivalsaari_2017}, and there is still a lack of a common set of approaches, models, and methodologies to facilitate engineering IoT-intensive software systems \cite{Zambonelli_2017}.
\item In practice, our initially developed principles together with the demonstrated techniques can help practitioners better implement IoT-native software projects. We argue that the three fundamental principles will stay valid and effective along the evolution of IoT, e.g., even after the fully-fledged software stacks are available on all the IoT devices.
\end{itemize}

The remainder of this paper is organized as follows. The three principles are introduced and discussed in Section
\ref{sec:PI}, \ref{sec:PII}, and \ref{sec:PIII} respectively.
Section \ref{sec:conclusion} draws conclusions and specifies our future work plan.



\section{Principle I: Just Enough}
\label{sec:PI}
The current big data business may incentivize collecting more data and holding them for longer periods of time \cite{Tene_Polonetsky_2013}. However, it is impractical and often unnecessary to centralize every bit of data from every single IoT device \cite{Kecskemeti_2017}.
Following this idea, we argue just enough collection, storage and processing of data when developing software systems in the IoT environments. 

\subsection{Minimization of Data Collection}
\label{subsec:datacollection}

The spirit of \textit{just enough} in this case naturally matches \textit{data minimization} that is a fundamental principle of the privacy law \cite{WhiteHouse_2012}. According to the privacy law, organizations must limit the collection of personal data to the minimum extent for their business needs. In addition to the concern about privacy, the limited resources and  battery life of IoT devices are the other critical reasons for software systems to reduce the data collection (e.g., lowering the sensing frequency).  

Besides the design-time data reduction and runtime data filtering \cite{Kecskemeti_2017}, we argue that on-device processing can further help minimize data collection, as it will avoid the centralization of raw data. Furthermore, the real-time data erasing should be recommended and preferred whenever it is possible during the data processing, because erasing the data at the earliest chance will also satisfy the minimization of data storage duration (cf.~Section \ref{subsec:storageDuration}). For example, the following techniques are particularly suitable for this case.

\begin{itemize}
\item \textbf{Running Statistics} can calculate the statistical indicators of the incoming data in a single pass. The recursive updates of statistical indicators and internal parameters are based on one data point at a time. Every data point can be erased immediately after being used. 

\item \textbf{Moving Window Statistics} only summarize the data within a predefined window, and keep updating the statistical summary along moving the window over the incoming data. Since the width of the window is fixed, each update of the statistical summary not only includes a new data point but also excludes the oldest one. A data point can be erased once it is excluded from the window.

\item \textbf{Streaming Algorithms} aim to cope with extremely large (or even infinite) sequential data in one pass (or a small number of passes) by using a constant size of space \cite{Navaz_Harous_2019}. By definition it is clear that the raw data streams are not supposed to be saved, and thus the streaming algorithms can be viewed as a superset of running statistics.

\end{itemize}

\subsection{Minimization of Data Storage Duration}
\label{subsec:storageDuration}

Unless the original data are critical for future verification or for cumulative data mining, it is generally unnecessary to keep them forever. This idea intuitively aligns with the decay theory, i.e. the stored information in human mind would gradually become unavailable for later retrieval as memory fades over time \cite{Ricker_2016}. The benefit of forgetting the previous information is that ``it lets us act in time, cognizant of, but not shackled by, past events. Through perfect memory we may lose a fundamental human capacity to live and act firmly in the present'' \cite{Ayalon_Toch_2013}. By analogy, to utilize storage resources more efficiently, we advocate to remove the data once they have satisfied the purposes for which they are collected. 

Considering the location and awareness of data consumption, we only discuss the data removal from the data sink's perspective. Note that we recognize two roles in a data flow, i.e.~data source and data sink. A data sink can in turn act as a data source when it delivers intermediate processing results to its downstream sinks.  

Furthermore, we distinguish between two patterns of data erasure according to the sink's controllability over the data's lifecycle.   

\begin{itemize}
\item \textbf{Controlled Data Erasure} is for the data that are consumed by the sink node only, and thus the erasure will purely depend on the data sink's decision. To minimize the data storage duration, the sink node may intentionally conduct data erasure right after its processing tasks are done. In particular, the received data can be discarded on the fly if they are processed in the streaming way (cf.~Section \ref{subsec:datacollection}).
\item \textbf{Data Self-Erasure} is for the data that are stored on the sink node while consumed by external stakeholders (e.g., the end users' review). In this case, the data erasure would have to wait for the uncertain external interventions. Driven by the concern about privacy, an alternative option is to let data be erased automatically based on some predefined rules. 
In fact, data self-erasure (a.k.a.~self-destruction) has become a successful innovation especially for the social media and communication applications, e.g., Snapchat\footnote{\url{https://support.snapchat.com/en-US/a/when-are-snaps-chats-deleted}} and WhatsApp\footnote{\url{https://faq.whatsapp.com/general/chats/about-disappearing-messages/}}.
\end{itemize}  


\subsection{Minimization of Data Processing Workloads}

To relieve the limitation of compute capability in the IoT environments, we can try to minimize the processing workloads from both the runtime and the design-time points of view. At runtime, a software system's data processing workloads would be proportional to the amount of user requests. Moreover, one user may issue the same request multiple times (e.g., due to the network interruption), and different users may request the same data simultaneously or at different times (e.g., in the multicast scenario). Therefore, in suitable cases, it is worth taking into account caching mechanisms to reduce the repeated data processing. A typical example that can demonstrate this strategy in the IoT domain is the mechanism \textit{content store} \cite{Amadeo_2018}.
Note that caching processing results does not violate the aforementioned minimization of the storage duration of raw data.  

At design time, the functional features of a software system could have essential impacts on the processing workloads. 
Therefore, we argue that practitioners should avoid over-designing IoT software systems and only implement just enough functional features to deal with real-world problems. This argument is confirmed by Elon Musk's recent talk about the five-step engineering process \cite{Pressman_2021}. In particular, (software) engineers are strongly suggested to clarify the requirements and to avoid the bias towards adding features when looking for a solution. A typical example that can demonstrate this argument in the IoT domain is: The researchers improved the performance of online skewness monitoring by skipping the explicit variance calculation  \cite{Li_Jhon_2021}. 

\section{Principle II: Just in Time}
\label{sec:PII}
By guaranteeing at least the just-in-time arrival of key data, we argue to tolerate certain processing delays to improve the communication efficiency or to relieve the network congestion within an IoT system. The ``just-in-time'' moments on different data sink nodes can be decided by referring to the plain system design and implementation. We use three different scenarios to explain this argument as follows.

The first and the most straightforward scenario is the one-off data transfer between a pair of source and sink nodes.
Assume transferring a data block $D$ takes time $T_D$ in the plain version of an IoT application. Alternatively, we can compress the data on the source node, transfer the compressed data, and then decompress the data on the sink node. If the alternative approach takes as much time as $T_D$ or less, the extra overhead of data compression and decompression will pay off, as it will at least save the bandwidth resource. In this case, we claim that the alternative approach satisfies the principle of \textit{just in time}, because the data arrival is guaranteed no later than its plain counterpart, as regulated in Eq.~(\ref{eq:compression}).  
\begin{equation}
\label{eq:compression}
  T_c + T_{D'} + T_{dec} \leqslant T_D
\end{equation}%
where $T_{D'}$ represents the latency of transferring the compressed data block $D'$, and $T_c$ and $T_{dec}$ indicate the time consumption of data compression and decompression respectively. 

The second scenario is the sequential transfer of a finite set of data points between a pair of source and sink nodes. Following the plain first-in-first-out (FIFO) fashion, we assume that transferring the first $k$ data points takes time $T_{\textit{fifo}(d_k)}$, i.e.~the wall-clock latency of transferring the $k$th data point $d_k$ is $T_{\textit{fifo}(d_k)}$. Inspired by the packet buffering feature of today's Internet protocols, we have developed application-level buffering mechanisms to improve the efficiency of data communication, as long as the arrival of the buffered key data satisfies Eq.~(\ref{eq:buffer}). The key data can either be a single data point that has high priority, or a group of data points that enjoys majority benefits. In both situations, we only need to select one data point as the reference to adjust the buffer size, and the reference's arrival is guaranteed no later than its FIFO counterpart. In practice, the dynamic buffer adjustment will require the prior knowledge about the IoT environment (via performance evaluation). 
\begin{equation}
\label{eq:buffer}
  T_{\textit{bf}(D)}+T_D \leqslant T_{\textit{fifo}(d_k)}, \quad\quad d_k \in D
\end{equation}%
where $T_{\textit{bf}(D)}$ stands for the time consumption of buffering a group of data points (including $d_k$) into a block $D$. In particular, $d_k$ acts as the reference that can bring majority benefits to $D$ without delaying its own arrival time. 

The third scenario is the data shuffling among multiple IoT devices. 
Data shuffling is pervasively required by the modern programming models and distributed computing frameworks (e.g., MapReduce and Dryad) \cite{Chowdhury_2011,Jin_Jia_2021}, and it may have a dominant effect on the completion time of a computing task. For example, the communication overhead for data shuffling can account for up to 70\% of the overall execution time of self-join applications \cite{Zhang_2013}. If a distributed workflow involves large amounts of shuffling events, the previous strategy of frequent data compression and decompression will be impractical and unacceptable. To address this challenge, we have introduced compact data structures (CDS) techniques \cite{Navarro_2016} to the development of data shuffling-intensive IoT applications. Compared with data compression, the CDS techniques can not only maintain data with less space, but they also enable the data to be queried, navigated, and manipulated directly in the compact form, i.e.~without being decompressed. By supplementing a CDS layer to the original data processing logic, we can still finish suitable tasks at least on time, with significantly less consumption of storage and network resources.     

Overall, this principle essentially promotes the value of preparation before making a job done, as stated by the idiom ``a beard well lathered is half shaved''. It should be noted that the increased processing workload here (for non-functional ``preparation'') does not violate the aforementioned sub-principle about minimizing processing workloads (for functional needs).

\section{Principle III: Just for ``Me''}
\label{sec:PIII}
Considering the distributed nature and limited resources of IoT devices, on-device processing would particularly be suitable for single-purpose and loosely coupled tasks, and the corresponding functional components would also follow the UNIX philosophy ``do one thing, and do it well''. Therefore, we argue that the architectural principle and style of microservices architecture (MSA) are still useful and valuable for building IoT-native software systems. For the purpose of conciseness, here we reuse the concept ``microservice'' to represent those IoT-native functional components, although the practical implementations may vary and not always fit in the de facto definition. For example, the popular microservice techniques and theoretical standards (e.g., containerization and self-maintained database) could be too heavy to implement on resource-constrained sensors and actuators, at least at this current stage.

Recall that the UNIX philosophy emphasizes high cohesion and loose coupling. Inspired by the idea about living with and exploiting the heterogeneity \cite{Morin_2017}, we further emphasize the awareness of environments. In particular, each microservice only needs to be aware of its own environment, which eventually leads to the principle \textit{just for ``me''}. As such, the IoT software development will be able to minimize the orchestration flows and mainly rely on the collaboration among microservices, via notifying and detecting their environments:

\begin{itemize}
\item \textbf{Notification} indicates the activity that a microservice updates the environment with its own status.
\item \textbf{Detection}  indicates the activity that a microservice requests the information of its environment.
\end{itemize}

\begin{figure}[!t]
  \centering
  \includegraphics[width=6.8cm,trim=60 645 325 110,clip]{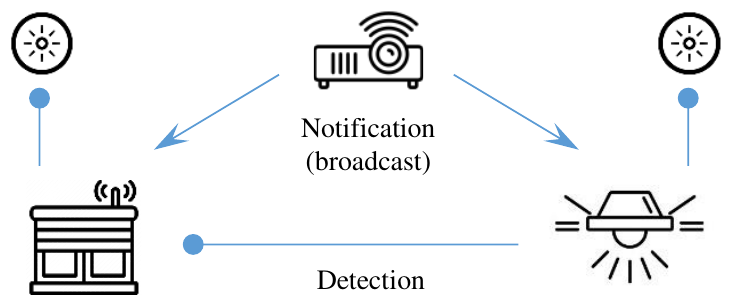}
  \caption{Microservice collaboration for room illumination.}
  \Description{Microservice collaboration for room illumination.}
\label{fig:collaboration}
\end{figure}

We employ the scenario of conference room illumination \cite{Zambonelli_2017} to demonstrate how to satisfy \textit{just for ``me''} in practice, as shown in \reffig{fig:collaboration}. 
For the convenience of description, we directly use the device names to represent the corresponding on-device microservices, 
 and we assume their business logic partnership in this demonstration as: The window shade needs outdoor brightness sensor (for deciding open or close), the ceiling light needs window shade and indoor brightness sensor (for deciding on or off), while the projector needs ceiling light and window shade (for adjusting the room illumination).  To clarify the environmental awareness scope, each microservice maintains a table of its predefined business logic partners, except for the standalone ones. Then:


(1) When the projector turns on for the first time, it broadcasts a semantic notification to all the devices within the same room. By matching the device names/interfaces, the ceiling lights and window shades will send their addresses as responses to the projector, and the projector will save these addresses into its partner table. After that, the projector will only broadcast future notifications to the saved addresses until it needs to update the partner table again.

(2) After sending responses to the projector and then interpreting the notification, the window shades will close (or remain closed) and the ceiling lights will turn off (or remain off).  

(3) Before the projector turns off, it will broadcast another notification to the relevant devices and verify their responses.

(4) After responding and interpreting the second notification, the window shades will open, unless it is dark night according to the outdoor brightness sensors. The ceiling lights will firstly detect the status of window shades. If the status is ``opening'', the ceiling lights will wait for a short period of time. Then, they further detect the natural daylight via indoor brightness sensors to decide whether they should turn on.

It is clear that compared to the study \cite{Zambonelli_2017}, our solution does not require any central lighting controller. Particularly, the semantic notifications are not specific control instructions. For example, the projector may notify its environment with ``I am turning on and expecting low illumination'', which does not physically guide what its business logic partners should do. Instead, the ceiling lights and window shades will be able to make different and appropriate actions, by interpreting the same notification into respective workflows. 

Overall, this principle essentially argues to further break the monolithic orchestration mechanism (e.g., cloud-centric coordination \cite{Truong_2015}), and to employ the choreography mechanism to compose individual microservices into an IoT application. 

\section{Conclusions and Future Work}
\label{sec:conclusion}

Since the term ``Internet of Things'' coined in 1999 \cite{Postscapes_2019}, the movement towards IoT has been happening and accelerating unprecedentedly in every sector of the world. In addition to the evolution of sensor technologies and IoT infrastructure, the emerging IoT applications have also become distinctive from the current mainstream application types. Given the trend in making everyday things programmable, we foresee the replacement of cloud-centric IoT solutions with IoT-native solutions that will implement software projects directly in the IoT environments.

Correspondingly, we argue the needs of dedicated efforts, techniques and theories for engineering IoT-native software systems. Unfortunately, the software engineering research in IoT is still at the early stage, not to mention the potential topic \textit{IoT-native software engineering}. Benefiting from attending a few various IoT projects, we particularly extracted experience and lessons from those software sections. This research summarizes our experience and lessons into three principles, and we claim them to be fundamental because (1) they were born from the unique IoT characteristics and (2) they match the IoT-native evolution direction. Based on the current effort, we plan to enlarge the collaborate scope and enrich the engineering principles to cover broader phases (ranging from system design to maintenance) of IoT-native software projects.


\bibliographystyle{ACM-Reference-Format}
\bibliography{sample-base}

\end{document}